\documentstyle[12pt,fleqn]{article}

\title{Cross Sections for the Electron Activation of \\
$\gamma$-Ray Fluorescence }

\author{Silviu Olariu\thanks{e-mail: olariu@ifin.nipne.ro} ~and Agata Olariu\\
National Institute of Physics and Nuclear Engineering, Tandem Laboratory\\
76900 Magurele, P.O. Box MG-6, Bucharest, Romania\\
\and
Yoshiaki Ito and Takeshi Mukoyama\\
Institute for Chemical Research, Kyoto University, Uji, Kyoto 611, Japan}

\begin{document}

\maketitle

\date{}

\abstract
  We report cross sections for the direct excitation of $\gamma$-ray 
transitions up to 200 keV by the transient electromagnetic fields 
of electrons from a beam, for electron incident kinetic energies of 500 keV 
and 5 MeV. The cross sections for the electron activation of $\gamma$-ray
fluorescence are of the order of 300 nb for an electron incident kinetic 
energy of 500 keV, and are of the order of 10 $\mu$b for an electron incident 
kinetic energy of 5 MeV. The electron excitation of nuclear transitions may 
lead to the development of pulsed sources of gamma radiation of narrowly 
defined energy.
\endabstract


\newpage

It has been shown \cite{1} that if a beam of incident electrons produces holes 
in the atomic shell of certain atomic species, this atomic excitation can be 
transferred to the nucleus. The nucleus thus raised in an excited state then 
decays by the emission of M\"{o}ssbauer gamma-ray photons. A pulse of incident 
electrons would produce in this way a pulse of M\"{o}ssbauer $\gamma$-ray 
photons, which have a narrowly-defined energy. The transfer of excitation 
from the atomic shell to the nucleus is due to the hyperfine interaction 
energy, and depends on the existence of a match 
between an atomic transition energy and a transition from the ground nuclear 
state to an excited nuclear state. The 
probability of the X-ray electron-nuclear transition (XENDT) is inversely 
proportional to the square of this detuning 
between the atomic transition energy and the nuclear transition energy.

In this work we study the {\it direct} excitation of $\gamma$-ray 
transitions up to 200 keV by the transient 
electromagnetic field of  electrons from an incident beam. This type of 
nuclear excitation is no longer restricted by
the requirement of the existence of a near resonance between the atomic and 
nuclear transition energies, and is in principle open to all nuclear species. 
Cross sections for the nuclear activation by electrons of $\gamma$-ray 
fluorescence have been calculated in this work using the results of 
Robl, \cite{2} who gave expressions of the cross sections in the Born 
approximation, for magnetic and electric multipole
transitions of arbitrary order. In these expressions, the nuclear properties 
are accounted for by the partial 
$\gamma$-ray widths of the excited M\"{o}ssbauer state. The partial widths 
are calculated in this work from tabulated 
nuclear data on half-lives, relative intensities and internal conversion 
coefficients.

The cross sections for the electron activation of $\gamma$-ray fluorescence
are of the order of 300 nb for incident electrons with a kinetic
energy of 500 keV, being about two orders of
magnitude lower than the cross sections for the excitation of
M\"{o}ssbauer transitions via electron-nuclear interactions, 
\cite{1} but the number of nuclei which can be excited directly by electrons 
is larger, because of the absence
of the near-resonance condition of \cite{1}. 
The cross sections for the electron activation of $\gamma$-ray fluorescence
are of the order of 10 $\mu$b for incident electrons with a kinetic
energy of 5 MeV.
The $\gamma$-ray activity induced by an electron beam incident on a sample 
containing the nuclear species under 
consideration is of the order of $10^4$ Bq/mA, for an electron incident 
kinetic energy of 500 keV. 
Much larger activities could be produced if the incident electron beam 
consists of {\it pulses} of electrons. Such pulsed sources would produce
$\gamma$-ray photons of narrowly defined energy. Because of their 
narrowly-defined energy, the $\gamma$-ray pulses produced in 
this way might be used for interference experiments, such as $\gamma$-ray 
holography. \cite{3} 
Pulsed sources of M\"{o}ssbauer radiation would be also
valuable for the cases when the half-life of a coventional M\"{o}ssbauer 
source is too short for practical use.

The cross section for the excitation by an incident electron of an 
$EL$ transition from a ground nuclear 
state $|g\rangle$ to an excited nuclear state $|e\rangle$ can be written 
in the Born approximation as \cite{2}
\begin{equation}
\sigma_{EL}=\frac{2I_e+1}{2I_g+1}\frac{\hbar\Gamma_{eg}^{EL}}{mc^2}F_{EL} ,
\end{equation}
where $\Gamma_{eg}^{EL}$ is the partial width for the emission of a 
$\gamma$-ray photon of multipolarity $EL$ 
in the transition $|e\rangle\rightarrow |g\rangle$,
$I_e, I_g$ are the spins of the excited and ground nuclear state, and 
$m$ is the electron mass. The quantity $F_{EL}$ 
is given by
\begin{eqnarray}
\lefteqn{F_{EL}=\frac{1}{L+1}\frac{\pi\alpha\hbar^2 c^2 m}
{E_\gamma^3 p_0^2}
\left[\frac{1}{4}(L-1)(E_0+E)^2K_{L-2}\right.}\nonumber\\
& &+\left(E_0 E-(L+1)m^2 c^4-\frac{3}{2}L (E_0^2+E^2)\right)K_{L-1}\nonumber\\
& & \left.+\left(2L E_0 E+\frac{1}{4}(7L+1) E_\gamma^2\right) K_L-\frac{1}{2} 
L E_\gamma^2 K_{L+1}\right] ,
\end{eqnarray}
where $\alpha$ is the fine structure constant, $E_\gamma$ is the energy of 
the $\gamma$-ray photon emitted in the transition $|e\rangle\rightarrow 
|g\rangle$, $E_0$ is the relativistic energy of the incident electron including
rest mass, 
$p_0$ is the momentum of the incident electron, 
$p_0 =(E_0^2-m^2c^4)^{1/2}/c$, and $E$ is the energy of the electron after 
excitation of the nucleus, $E=E_0-E_\gamma$.
The dimensionless quantities $K_L$ are defined by the recurrence relation
\begin{equation}
K_{L+2}-2K_{L+1}+K_L=\frac{1}{L+1}\left(x_2^{L+1}-x_1^{L+1}\right) ,
\end{equation}
with
\begin{equation}
K_0=\frac{p_0p}{m^2c^2}, \;
K_1=K_0+2\ln\frac{mc^2E_\gamma}{E_0 E-m^2c^4-p_0pc^2} ,
\end{equation}
where $p$ is the momentum of the electron after excitation of the nuclear 
state, $p =(E^2-m^2c^4)^{1/2}/c$, and the dimensionless quantities $x_1, x_2$
are 
\begin{equation}
x_1=\frac{(p_0-p)^2}{\hbar^2 k^2},\;x_2=\frac{(p_0+p)^2}{\hbar^2 k^2}.
\end{equation}

The cross section for the excitation by an incident electron of an 
$ML$ transition from a ground nuclear 
state $|g\rangle$ to an excited nuclear state $|e\rangle$ can be written 
in the Born approximation as \cite{2}
\begin{equation}
\sigma_{ML}=\frac{2I_e+1}{2I_g+1}\frac{\hbar\Gamma_{eg}^{ML}}{mc^2}F_{ML} ,
\end{equation}
where $\Gamma_{eg}^{ML}$ is the partial width for the emission of a 
$\gamma$-ray photon of multipolarity $ML$ 
in the transition $|e\rangle\rightarrow |g\rangle$,
and the quantity $F_{ML}$ is given by
\begin{equation}
F_{ML}=\frac{\pi\alpha\hbar^2 c^2 m}{E_\gamma^3p_0^2}
\left[(E_0 E-m^2 c^4)K_L
+\frac{1}{4}E_\gamma^2(K_{L+1}-x_1 x_2 K_{L-1})\right] .
\end{equation}

The partial $\gamma$-ray width $\Gamma_{eg}$ of the transition 
$|e\rangle\rightarrow |g\rangle$ can be calculated as
\begin{equation}
\Gamma_{eg}=\frac{\ln 2}{t_e}\frac{R_{eg}}{\sum_l (1+\alpha_{el}R_{el})} ,
\end{equation}
where $t_e$ is the half-life of the excited state $|e\rangle$, the index 
$l$ designates all nuclear states lower than $|e\rangle$, including the 
ground state $|g\rangle$, $R_{el}$ is
the relative intensity of the transition $|e\rangle\rightarrow |g\rangle$, 
and $\alpha_{el}$ are the internal conversion coefficients.
Among the cases studied in this work, we encountered mixed transitions only of
the $M1+E2$ type. In the case of mixed $M1+E2$ transitions, the internal
conversion coefficients have been calculated as 
\begin{equation}
\alpha=\frac{1}{1+\delta^2}\alpha_{M1}+\frac{\delta^2}{1+\delta^2}\alpha_{E2},
\end{equation}
the partial widths are
\begin{equation}
\Gamma_{eg}^{M1}=\frac{1}{1+\delta^2}\Gamma_{eg},\;
\Gamma_{eg}^{E2}=\frac{\delta^2}{1+\delta^2}\Gamma_{eg},
\end{equation}
and the total cross section for the electron excitation of the $|e\rangle$
state is
\begin{equation}
\sigma_{M1+E2}=\sigma_{M1}+\sigma_{E2},
\end{equation}
where $\delta$ is the mixing ratio for the $M1$ and $E2$  components
of the $\gamma$-ray transition $|e\rangle\rightarrow |g\rangle$. \cite{4b}
A pure $M1$ transition corresponds to $\delta=0$, and a pure $E2$ 
corresponds formally to $\delta=\infty$.
For the $E1, M2$ and $E3$ transitions encountered in this work, we had
$\Gamma_{eg}^{E1}=\Gamma_{eg}$, 
$\Gamma_{eg}^{M2}=\Gamma_{eg}$, and $\Gamma_{eg}^{E3}=\Gamma_{eg}$.

The cross section $\sigma_\gamma$ for the emission 
of a $\gamma$-ray photon to be induced by an incident electron is lower than 
$\sigma_{EL}$, $\sigma_{ML}$ or $\sigma_{M1+E2}$
by the factor $f=\Gamma_{eg}/\Gamma_e$, where $\Gamma_e=\ln 2/t_e$,
\begin{equation}
\sigma_\gamma=f\sigma_e ,
\end{equation}
where $\sigma_e$ denotes $\sigma_{EL}$, $\sigma_{ML}$
or $\sigma_{M1+E2}$, according to the case.
In the case of a M\"{o}ssbauer experiment, the cross section $\sigma_\gamma$ 
must be multiplied by the recoil-free fraction.

The main limitation of the previous expressions for the electron 
excitation cross section arises from the condition of
validity of the Born approximation, which requires that 
$Z\alpha(E/pc)\ll 1$, where $Z$ is the proton number of the 
nucleus. As discussed by Robl, \cite{2} there is also an upper limitation 
of the photon energy $E_\gamma$ of  
about 10 MeV which is not of concern for the present work, because we are 
studying transitions of much lower energy.

We have calculated cross sections for 144 cases of exitation of nuclear 
states having energies less than 200 keV, connected 
to the ground state by E1, M1, E2, M2 or E3 transitions, for which the 
half-lives $t_e$, the relative intensities
$R_{el}$ and the mixing ratios were known. \cite{4} We have calculated the
internal conversion coefficients by interpolation \cite{5},
\cite{6} and have also used values calculated on-line at the National 
Nuclear Data Center, Brookhaven. \cite{7}
The cross section $\sigma_e$ and $\sigma_\gamma$ are listed in Table I, for
a kinetic energy of the incident electron of $E_0-mc^2=500$ keV,
and $E_0-mc^2=5$ MeV. 
It can be seen from Table I that the largest cross section $\sigma_\gamma$
for an electron kinetic energy of 500 keV for the
emission of a $\gamma$-ray photon has the value 377 nb. The cross 
sections at this electron energy are about two orders of magnitude lower than
the cross sections for the excitation of M\"{o}ssbauer transitions via 
electron-nuclear interactions, 
\cite{1} but the number of nuclear states which can be excited directly by 
electrons is larger, because of the absence
of the near-resonance condition of \cite{1}. 
It can be seen from Table I that the cross section for an incident kinetic
energy of the electron of 5 MeV are significantly larger than the cross 
sections corresponding for a kinetic energy of 500 keV.
We have also given in Table I values of the quantity $Z\alpha E/pc$.

As estimated in \cite{1}, when the incident electron beam is incident 
on a suitable target, such cross sections 
result in activities of the order of $10^4$ Bq/mA for an electron incident
kinetic energy of 500 keV. 
In the case of a {\it pulsed} incident electron beam the resulting activities
would be much larger
over the period of the pulse, because of the much larger incident
current. The photons emitted by such a pulsed 
$\gamma$-ray source would have a narrowly defined energy, and would be 
suitable for interference experiments, such
as holography. \cite{3} 

Electron activated $\gamma$-ray sources could also be of interest for 
M\"{o}ssbauer experiments in cases where the 
half-life of conventional radioactive sources is too short to be practical. 
For example, among the cases listed in Table I, the 
radioactive M\"{o}ssbauer sources which could be used for 
$^{162}$Er, $^{188}$Os, $^{164}$Er, $^{165}$Ho, $^{184}$Os, $^{180}$W,
$^{187}$Re, $^{180}$Hf, $^{162}$Dy, $^{174}$Hf, $^{176}$Yb, $^{187}$Os,
$^{160}$Gd, $^{164}$Dy, $^{163}$Dy
have a half-life less than 1 day, and no source is
available for $^{40}$K. 

Although the cross sections of X-ray electron-nuclear double transitions 
discussed in \cite{1} and the cross sections
for the direct excitation of nuclear transitions are not very large, 
these processes appear to be of real interest 
for the development of electron-activated sources of $\gamma$-ray 
photons of narrowly defined energy.\\

{\it Acknowledgment} One of the authors (S.O.) acknowledges the 
financial support of the University of Kyoto during the preparation 
of this work.

\newpage

TABLE CAPTION \\

TABLE I. Cross section $\sigma_e$ for the electron excitation of nuclear
states and cross section $\sigma_\gamma$ for the generation of fluorescence
$\gamma$-ray 
photons of energy $E_\gamma$ by electrons with incident kinetic energy 
$E_0-mc^2=500$ keV, and $E_0-mc^2=5$ MeV. The powers of 10 are represented 
with the aid of the symbol E, so that the value $3\times 10^{-10}$ is written
as 3E-10.

\newpage

\setlength{\oddsidemargin}{-2.5cm}
\setlength{\topmargin}{-2cm}
\setlength{\textwidth}{19cm}
\setlength{\textheight}{24.5cm}

\scriptsize

\begin{tabular}{ccccccccccccc}
\hline
nucleus	&      nat.ab.	&	$E_\gamma,$ &	$t_e$,   &	multipole&
$\delta$& $f$  & $\sigma_e$, nb &$\sigma_\gamma$, nb   &$Z\alpha E/pc$, 
             & $\sigma_e$, nb &$\sigma_\gamma$, nb   &$Z\alpha E/pc$, \\
        &        \%     &       keV         &    s      
&       $|e\rangle\rightarrow |g\rangle$   &      
        &    & 500 keV           & 500 keV       &  500 keV       
        &  5 MeV           & 5 MeV       &  5 MeV       \\
\hline
$^{ 169}$Tm & 100.00 & 130.5 & 3E-10 & E2 &  & 3.19E-1 & 1.18E3 & 3.77E2 
& 0.618 & 3.54E4 & 1.13E4 & 0.506 \\
$^{ 156}$Dy & 0.06 & 137.8 & 8.2E-10 & E2 &  & 5.39E-1 & 6.87E2 & 3.7E2 
& 0.594 & 2.09E4 & 1.12E4 & 0.484 \\
$^{ 154}$Gd & 2.18 & 123.1 & 1.2E-9 & E2 &  & 4.55E-1 & 7.33E2 & 3.33E2 
& 0.571 & 2.17E4 & 9.85E3 & 0.469 \\
$^{ 162}$Er & 0.14 & 102.0 & 1.2E-9 & E2 &  & 2.65E-1 & 1.16E3 & 3.08E2 
& 0.600 & 3.3E4 & 8.75E3 & 0.498 \\
$^{ 152}$Sm & 26.70 & 121.8 & 1.4E-9 & E2 &  & 4.61E-1 & 6.52E2 & 3.01E2 
& 0.553 & 1.92E4 & 8.87E3 & 0.454 \\
$^{ 150}$Nd & 5.64 & 130.2 & 1.4E-9 & E2 &  & 5.36E-1 & 5.29E2 & 2.83E2 
& 0.538 & 1.58E4 & 8.49E3 & 0.440 \\
$^{ 190}$Os & 26.40 & 186.7 & 3.6E-10 & E2 &  & 7.01E-1 & 3.98E2 & 2.79E2 
& 0.707 & 1.33E4 & 9.32E3 & 0.557 \\
$^{ 186}$Os & 1.58 & 137.2 & 8.2E-10 & E2 &  & 4.35E-1 & 5.74E2 & 2.5E2 
& 0.684 & 1.74E4 & 7.57E3 & 0.557 \\
$^{ 158}$Dy & 0.10 & 98.9 & 1.7E-9 & E2 &  & 2.59E-1 & 9.31E2 & 2.41E2 
& 0.581 & 2.64E4 & 6.83E3 & 0.484 \\
$^{ 186}$W & 28.60 & 122.6 & 1E-9 & E2 &  & 3.58E-1 & 6.74E2 & 2.41E2 
& 0.660 & 1.99E4 & 7.12E3 & 0.542 \\
$^{ 188}$Os & 13.30 & 155.0 & 7.1E-10 & E2 &  & 5.49E-1 & 4.34E2 & 2.38E2 
& 0.691 & 1.36E4 & 7.46E3 & 0.557 \\
$^{ 164}$Er & 1.61 & 91.4 & 1.5E-9 & E2 &  & 1.91E-1 & 1.18E3 & 2.26E2 
& 0.597 & 3.29E4 & 6.3E3 & 0.498 \\
$^{ 165}$Ho & 100.00 & 94.7 & 2.2E-11 & M1+E2 & 0.155 & 2.4E-1 & 8.53E2 
& 2.05E2 & 0.589 & 1.44E4 & 3.46E3 & 0.491 \\
$^{ 184}$W & 30.67 & 111.2 & 1.3E-9 & E2 &  & 2.76E-1 & 7.18E2 & 1.98E2 
& 0.656 & 2.08E4 & 5.73E3 & 0.542 \\
$^{ 156}$Gd & 20.47 & 89.0 & 2.2E-9 & E2 &  & 2.03E-1 & 9.55E2 & 1.94E2 
& 0.561 & 2.66E4 & 5.38E3 & 0.469 \\
$^{ 185}$Re & 37.40 & 125.4 & 1E-11 & M1+E2 & 0.18 & 2.63E-1 & 7.21E2 
& 1.9E2 & 0.670 & 1.2E4 & 3.17E3 & 0.550 \\
$^{ 184}$Os & 0.02 & 119.8 & 1.2E-9 & E2 &  & 3.16E-1 & 5.88E2 & 1.86E2 
& 0.677 & 1.73E4 & 5.46E3 & 0.557 \\
$^{ 180}$W & 0.13 & 103.6 & 1.3E-9 & E2 &  & 2.24E-1 & 8.26E2 & 1.85E2 
& 0.653 & 2.36E4 & 5.28E3 & 0.542 \\
$^{ 168}$Yb & 0.13 & 87.7 & 1.5E-9 & E2 &  & 1.55E-1 & 1.18E3 & 1.82E2 
& 0.613 & 3.27E4 & 5.06E3 & 0.513 \\
$^{ 160}$Dy & 2.34 & 86.8 & 2E-9 & E2 &  & 1.76E-1 & 1.03E3 & 1.8E2 
& 0.578 & 2.85E4 & 5E3 & 0.484 \\
$^{ 182}$W & 26.30 & 100.1 & 1.4E-9 & E2 &  & 2.02E-1 & 8.31E2 & 1.68E2 
& 0.652 & 2.36E4 & 4.76E3 & 0.542 \\
$^{ 187}$Re & 62.60 & 134.2 & 1.1E-11 & M1+E2 & 0.175 & 3.03E-1 & 5.53E2 
& 1.67E2 & 0.674 & 8.74E3 & 2.64E3 & 0.550 \\
$^{ 178}$Hf & 27.30 & 93.2 & 1.5E-9 & E2 &  & 1.74E-1 & 9.6E2 & 1.67E2 
& 0.633 & 2.69E4 & 4.67E3 & 0.528 \\
$^{ 180}$Hf & 35.10 & 93.3 & 1.5E-9 & E2 &  & 1.75E-1 & 9.44E2 & 1.65E2 
& 0.633 & 2.65E4 & 4.62E3 & 0.528 \\
$^{ 181}$Ta & 99.99 & 136.3 & 3.9E-11 & M1+E2 & 0.41 & 3.6E-1 & 4.41E2 
& 1.59E2 & 0.656 & 1.13E4 & 4.06E3 & 0.535 \\
$^{ 176}$Hf & 5.21 & 88.3 & 1.4E-9 & E2 &  & 1.45E-1 & 1.09E3 & 1.58E2 
& 0.631 & 3.04E4 & 4.4E3 & 0.528 \\
$^{ 170}$Yb & 3.05 & 84.3 & 1.6E-9 & E2 &  & 1.35E-1 & 1.16E3 & 1.57E2 
& 0.612 & 3.21E4 & 4.33E3 & 0.513 \\
$^{ 166}$Er & 33.60 & 80.6 & 1.8E-9 & E2 &  & 1.27E-1 & 1.21E3 & 1.53E2 
& 0.594 & 3.31E4 & 4.19E3 & 0.498 \\
$^{ 162}$Dy & 25.50 & 80.7 & 2.2E-9 & E2 &  & 1.38E-1 & 1.09E3 & 1.5E2 
& 0.576 & 2.98E4 & 4.12E3 & 0.484 \\
$^{ 158}$Gd & 24.84 & 79.5 & 2.5E-9 & E2 &  & 1.43E-1 & 1.05E3 & 1.5E2 
& 0.559 & 2.88E4 & 4.11E3 & 0.469 \\
$^{ 154}$Sm & 22.70 & 82.0 & 3E-9 & E2 &  & 1.69E-1 & 8.88E2 & 1.5E2 
& 0.542 & 2.44E4 & 4.12E3 & 0.454 \\
$^{ 40}$K & 0.01 & 29.8 & 4.2E-9 & (M1) &  & 7.75E-1 & 1.91E2 & 1.48E2 
& 0.162 & 3.59E2 & 2.78E2 & 0.139 \\
$^{ 168}$Er & 26.80 & 79.8 & 1.9E-9 & E2 &  & 1.22E-1 & 1.19E3 & 1.45E2 
& 0.594 & 3.25E4 & 3.98E3 & 0.498 \\
$^{ 174}$Hf & 0.16 & 91.0 & 1.7E-9 & E2 &  & 1.6E-1 & 8.94E2 & 1.43E2 
& 0.632 & 2.49E4 & 4E3 & 0.528 \\
$^{ 170}$Er & 14.90 & 78.6 & 1.9E-9 & E2 &  & 1.16E-1 & 1.21E3 & 1.4E2 
& 0.593 & 3.31E4 & 3.83E3 & 0.498 \\
$^{ 176}$Yb & 12.70 & 82.1 & 1.8E-9 & E2 &  & 1.24E-1 & 1.11E3 & 1.37E2 
& 0.612 & 3.04E4 & 3.76E3 & 0.513 \\
$^{ 187}$Os & 1.60 & 74.3 & 2E-11 & M1+E2 & 0.08 & 1.18E-1 & 1.15E3 & 1.35E2 
& 0.662 & 1.24E4 & 1.46E3 & 0.557 \\
$^{ 172}$Yb & 21.90 & 78.7 & 1.6E-9 & E2 &  & 1.06E-1 & 1.26E3 & 1.33E2 
& 0.611 & 3.43E4 & 3.64E3 & 0.513 \\
$^{ 160}$Gd & 21.86 & 75.3 & 2.7E-9 & E2 &  & 1.18E-1 & 1.09E3 & 1.28E2 
& 0.557 & 2.95E4 & 3.49E3 & 0.469 \\
$^{ 164}$Dy & 28.20 & 73.4 & 2.4E-9 & E2 &  & 9.93E-2 & 1.17E3 & 1.16E2 
& 0.574 & 3.16E4 & 3.14E3 & 0.484 \\
$^{ 175}$Lu & 97.41 & 113.8 & 9.9E-11 & M1+E2 & 0.464 & 2.82E-1 & 4.07E2 
& 1.15E2 & 0.630 & 1.07E4 & 3.02E3 & 0.520 \\
$^{ 177}$Hf & 18.61 & 112.9 & 5.8E-10 & M1+E2 & -4.7 & 3.05E-1 & 3.76E2 
& 1.15E2 & 0.639 & 1.09E4 & 3.33E3 & 0.528 \\
$^{ 174}$Yb & 31.80 & 76.5 & 1.8E-9 & E2 &  & 9.5E-2 & 1.21E3 & 1.15E2 
& 0.610 & 3.29E4 & 3.12E3 & 0.513 \\
$^{ 179}$Hf & 13.63 & 122.8 & 3.7E-11 & M1+E2 & -0.27 & 3.06E-1 & 3.64E2 
& 1.12E2 & 0.642 & 7.97E3 & 2.44E3 & 0.528 \\
$^{ 173}$Yb & 16.12 & 78.6 & 4.6E-11 & M1+E2 & -0.224 & 1.24E-1 & 8.55E2 
& 1.06E2 & 0.611 & 1.88E4 & 2.33E3 & 0.513 \\
$^{ 103}$Rh & 100.00 & 53.3 & 1.1E-9 & M1 &  & 3.21E-1 & 2.86E2 & 9.19E1 
& 0.388 & 5.93E2 & 1.9E2 & 0.330 \\
$^{ 153}$Eu & 52.20 & 83.4 & 7.9E-10 & M1+E2 & 0.8 & 2.07E-1 & 4.07E2 
& 8.42E1 & 0.551 & 1.1E4 & 2.27E3 & 0.462 \\
$^{ 153}$Eu & 52.20 & 97.4 & 2E-10 & E1 &  & 7.62E-1 & 1.05E2 & 8.01E1 
& 0.555 & 2.02E2 & 1.54E2 & 0.462 \\
\hline
\end{tabular}

\begin{tabular}{ccccccccccccc}
\hline
nucleus	&      nat.ab.	&	$E_\gamma,$ &	$t_e$,   &	multipole&
$\delta$& $f$  & $\sigma_e$, nb &$\sigma_\gamma$, nb   &$Z\alpha E/pc$, 
             & $\sigma_e$, nb &$\sigma_\gamma$, nb   &$Z\alpha E/pc$, \\
        &        \%     &       keV         &    s      &       
$|e\rangle\rightarrow |g\rangle$   &      
        &    & 500 keV           & 500 keV       &  500 keV       
        &  5 MeV           & 5 MeV       &  5 MeV       \\
\hline
$^{ 183}$W & 14.30 & 46.5 & 1.9E-10 & M1+E2 & -0.081 & 1.03E-1 & 7.57E2 
& 7.8E1 & 0.637 & 1.15E4 & 1.18E3 & 0.542 \\
$^{ 159}$Tb & 100.00 & 58.0 & 5.4E-11 & M1+E2 & 0.119 & 8.22E-2 & 9.43E2 
& 7.75E1 & 0.562 & 1.67E4 & 1.37E3 & 0.476 \\
$^{ 163}$Dy & 24.90 & 167.3 & 3.4E-10 & (E2) &  & 4.56E-1 & 1.67E2 & 7.59E1 
& 0.605 & 5.35E3 & 2.44E3 & 0.484 \\
$^{ 157}$Gd & 15.65 & 54.5 & 1.3E-10 & M1+E2 & 0.19 & 7.46E-2 & 9.26E2 
& 6.91E1 & 0.552 & 2.06E4 & 1.54E3 & 0.469 \\
$^{ 153}$Eu & 52.20 & 193.1 & 2E-10 & E2(calc)) &  & 4.77E-1 & 1.34E2 
& 6.38E1 & 0.589 & 4.52E3 & 2.16E3 & 0.462 \\
$^{ 55}$Mn & 100.00 & 126.0 & 2.6E-10 & M1(+E2) & 0.052 & 9.85E-1 & 5.64E1 
& 5.56E1 & 0.223 & 2.75E2 & 2.71E2 & 0.183 \\
$^{ 155}$Gd & 14.80 & 60.0 & 1.9E-10 & M1+E2 & -0.197 & 9.59E-2 & 5.36E2 
& 5.13E1 & 0.554 & 1.19E4 & 1.14E3 & 0.469 \\
$^{ 163}$Dy & 24.90 & 73.4 & 1.5E-9 & E2 &  & 9.95E-2 & 4.93E2 & 4.9E1 
& 0.574 & 1.33E4 & 1.33E3 & 0.484 \\
$^{ 171}$Yb & 14.30 & 66.7 & 8.1E-10 & M1+E2 & -0.705 & 7.74E-2 & 5.95E2 
& 4.6E1 & 0.607 & 1.56E4 & 1.21E3 & 0.513 \\
$^{ 167}$Er & 22.95 & 79.3 & 1.1E-10 & M1+E2 & -0.2 & 1.47E-1 & 3.1E2 
& 4.54E1 & 0.593 & 6.5E3 & 9.52E2 & 0.498 \\
$^{ 187}$Os & 1.60 & 187.4 & 1E-10 & E2 &  & 1.9E-1 & 2.29E2 & 4.35E1 
& 0.707 & 7.64E3 & 1.45E3 & 0.557 \\
$^{ 193}$Ir & 62.70 & 139.0 & 8.2E-11 & M1+E2 & -0.329 & 2.95E-1 & 1.41E2 
& 4.16E1 & 0.693 & 3.32E3 & 9.8E2 & 0.564 \\
$^{ 191}$Ir & 37.30 & 129.4 & 1.2E-10 & M1+E2 & -0.4 & 2.57E-1 & 1.52E2 
& 3.89E1 & 0.690 & 3.85E3 & 9.88E2 & 0.564 \\
$^{ 171}$Yb & 14.30 & 75.9 & 1.6E-9 & E2 &  & 6.56E-2 & 5.69E2 & 3.73E1 
& 0.610 & 1.55E4 & 1.02E3 & 0.513 \\
$^{ 183}$W & 14.30 & 99.1 & 7.7E-10 & E2 &  & 8.84E-2 & 4.1E2 & 3.62E1 
& 0.652 & 1.16E4 & 1.03E3 & 0.542 \\
$^{ 199}$Hg & 16.87 & 158.4 & 2.5E-9 & E2 &  & 5.22E-1 & 6.4E1 & 3.34E1 
& 0.729 & 2.02E3 & 1.05E3 & 0.586 \\
$^{ 107}$Ag & 51.84 & 32.5 & 2.9E-9 & M1+E2 & 0.074 & 7.98E-2 & 4.18E2 
& 3.33E1 & 0.402 & 7.23E3 & 5.77E2 & 0.344 \\
$^{ 47}$Ti & 7.30 & 159.4 & 2.1E-10 & M1+E2 & -0.099 & 9.94E-1 & 3.31E1 
& 3.29E1 & 0.201 & 2.68E2 & 2.66E2 & 0.161 \\
$^{ 19}$F & 100.00 & 109.9 & 5.9E-10 & E1(calc) &  & 1 & 3.05E1 & 3.05E1 
& 0.080 & 6.13E1 & 6.13E1 & 0.066 \\
$^{ 117}$Sn & 7.68 & 158.6 & 2.8E-10 & M1+E2 & 0.0133 & 8.63E-1 & 2.75E1 
& 2.37E1 & 0.456 & 8.42E1 & 7.27E1 & 0.367 \\
$^{ 61}$Ni & 1.14 & 67.4 & 5.3E-9 & M1+E2 & 0.0076 & 8.9E-1 & 2.26E1 
& 2.01E1 & 0.243 & 5.23E1 & 4.66E1 & 0.205 \\
$^{ 169}$Tm & 100.00 & 118.2 & 6.2E-11 & E2 &  & 2.87E-2 & 6.56E2 & 1.88E1 
& 0.614 & 1.92E4 & 5.51E2 & 0.506 \\
$^{ 189}$Os & 16.10 & 69.5 & 1.6E-9 & M1+E2 & 0.69 & 8.67E-2 & 1.97E2 
& 1.71E1 & 0.660 & 5.18E3 & 4.49E2 & 0.557 \\
$^{ 155}$Gd & 14.80 & 146.1 & 1E-10 & E2 &  & 7.18E-2 & 2.19E2 & 1.57E1 
& 0.579 & 6.76E3 & 4.85E2 & 0.469 \\
$^{ 157}$Gd & 15.65 & 131.4 & 9.5E-11 & E2 &  & 4.94E-2 & 2.81E2 & 1.39E1 
& 0.574 & 8.42E3 & 4.16E2 & 0.469 \\
$^{ 101}$Ru & 17.00 & 127.2 & 6.6E-10 & M1+E2 & 0.148 & 8.55E-1 & 1.44E1 
& 1.23E1 & 0.394 & 2.01E2 & 1.71E2 & 0.323 \\
$^{ 147}$Sm & 15.00 & 121.2 & 8E-10 & M1+E2 & -0.33 & 4.95E-1 & 2.38E1 
& 1.18E1 & 0.553 & 5.69E2 & 2.82E2 & 0.454 \\
$^{ 131}$Xe & 21.20 & 80.2 & 4.8E-10 & M1+E2 & 0.0165 & 4.34E-1 & 2.29E1 
& 9.95 & 0.471 & 6.33E1 & 2.75E1 & 0.396 \\
$^{ 235}$U & 0.72 & 46.2 & 6E-11 & M1+E2 & 0.14 & 1.72E-2 & 5.26E2 & 9.05 
& 0.791 & 1.09E4 & 1.88E2 & 0.674 \\
$^{ 127}$I & 100.00 & 57.6 & 2E-9 & M1+E2 & -0.084 & 2.08E-1 & 4.09E1 
& 8.51 & 0.458 & 5.57E2 & 1.16E2 & 0.388 \\
$^{ 173}$Yb & 16.12 & 179.4 & 3.2E-11 & E2 &  & 5.62E-2 & 1.5E2 & 8.44 
& 0.647 & 4.94E3 & 2.77E2 & 0.513 \\
$^{ 119}$Sn & 8.59 & 23.9 & 1.8E-8 & M1(+E2) &  & 1.61E-1 & 4.96E1 & 7.96 
& 0.426 & 9.03E1 & 1.45E1 & 0.366 \\
$^{ 161}$Dy & 18.90 & 25.7 & 2.9E-8 & E1 &  & 3.01E-1 & 2.6E1 & 7.82 
& 0.563 & 3.72E1 & 1.12E1 & 0.484 \\
$^{ 167}$Er & 22.95 & 178.0 & 5.5E-11 & E2(calc) &  & 7.25E-2 & 1.06E2 
& 7.66 & 0.628 & 3.47E3 & 2.51E2 & 0.499 \\
$^{ 195}$Pt & 33.80 & 98.8 & 1.7E-10 & M1+E2 & -0.13 & 1.22E-1 & 6.26E1 
& 7.62 & 0.687 & 8.93E2 & 1.09E2 & 0.572 \\
$^{ 99}$Ru & 12.70 & 89.7 & 2E-8 & E2+M1 & -1.56 & 4E-1 & 1.86E1 & 7.43 
& 0.386 & 5.14E2 & 2.06E2 & 0.323 \\
$^{ 129}$Xe & 26.40 & 39.6 & 9.7E-10 & M1+E2 & -0.027 & 7.43E-2 & 9.62E1 
& 7.15 & 0.463 & 5.37E2 & 3.99E1 & 0.396 \\
$^{ 169}$Tm & 100.00 & 8.4 & 4.1E-9 & M1+E2 & 0.033 & 5.62E-3 & 1.19E3 
& 6.71 & 0.585 & 2.4E4 & 1.35E2 & 0.506 \\
$^{ 147}$Sm & 15.00 & 197.3 & 1.2E-9 & E2 &  & 7.07E-1 & 8.7 & 6.16 & 0.581 
& 2.97E2 & 2.1E2 & 0.455 \\
$^{ 125}$Te & 7.14 & 35.5 & 1.5E-9 & M1+E2 & 0.029 & 6.65E-2 & 8.57E1 & 5.7 
& 0.445 & 5.72E2 & 3.8E1 & 0.381 \\
$^{ 159}$Tb & 100.00 & 137.5 & 4.1E-11 & E2(calc) &  & 2.33E-2 & 2.39E2 
& 5.58 & 0.585 & 7.26E3 & 1.69E2 & 0.476 \\
$^{ 232}$Th & 100.00 & 49.4 & 3.5E-10 & E2 &  & 2.97E-3 & 1.85E3 & 5.49 
& 0.775 & 4.8E4 & 1.43E2 & 0.660 \\
$^{ 85}$Rb & 72.17 & 151.2 & 7.1E-10 & M1+E2 & 0.072 & 9.53E-1 & 5.3 & 5.05 
& 0.336 & 3.2E1 & 3.05E1 & 0.271 \\
$^{ 197}$Au & 100.00 & 77.4 & 1.9E-9 & M1+E2 & -0.368 & 1.84E-1 & 2.71E1 
& 4.99 & 0.689 & 6.79E2 & 1.25E2 & 0.579 \\
$^{ 201}$Hg & 13.18 & 167.4 & 2.6E-11 & M1+E2 & 0.08 & 2.43E-1 & 1.9E1 
& 4.63 & 0.734 & 1.22E2 & 2.97E1 & 0.586 \\
$^{ 155}$Gd & 14.80 & 86.5 & 6.5E-9 & E1 &  & 6.84E-1 & 6.6 & 4.51 & 0.560 
& 1.22E1 & 8.36 & 0.469 \\
$^{ 238}$U & 99.27 & 44.9 & 2E-10 & E2 &  & 1.6E-3 & 2.73E3 & 4.36 & 0.791 
& 7.04E4 & 1.12E2 & 0.674 \\
$^{ 123}$Sb & 42.64 & 160.3 & 6.1E-10 & M1+E2 & 0.063 & 8.55E-1 & 4.84 
& 4.14 & 0.466 & 2.56E1 & 2.18E1 & 0.374 \\
\hline
\end{tabular}

\begin{tabular}{ccccccccccccc}
\hline
nucleus	&      nat.ab.	&	$E_\gamma,$ &	$t_e$,   &	multipole&
$\delta$& $f$  & $\sigma_e$, nb &$\sigma_\gamma$, nb   &$Z\alpha E/pc$, 
             & $\sigma_e$, nb &$\sigma_\gamma$, nb   &$Z\alpha E/pc$, \\
        &        \%     &       keV         &    s      &       
$|e\rangle\rightarrow |g\rangle$   &      
        &    & 500 keV           & 500 keV       &  500 keV       
        &  5 MeV           & 5 MeV       &  5 MeV       \\
\hline
$^{ 155}$Gd & 14.80 & 105.3 & 1.2E-9 & E1 &  & 4.52E-1 & 8.16 & 3.69 
& 0.566 & 1.62E1 & 7.3 & 0.469 \\
$^{ 145}$Nd & 8.30 & 72.5 & 7.2E-10 & M1 &  & 2.14E-1 & 1.56E1 & 3.35 
& 0.522 & 3.47E1 & 7.43 & 0.440 \\
$^{ 57}$Fe & 2.20 & 14.4 & 9.8E-8 & M1+E2 & 0.00219 & 1.05E-1 & 3.05E1 
& 3.19 & 0.221 & 5.76E1 & 6.02 & 0.191 \\
$^{ 187}$Os & 1.60 & 75.0 & 2.2E-9 & E2 &  & 2.13E-2 & 1.48E2 & 3.16 
& 0.662 & 4.03E3 & 8.58E1 & 0.557 \\
$^{ 67}$Zn & 4.10 & 184.6 & 1E-9 & M1+E2 & 0.34 & 8.51E-1 & 3.71 & 3.16 
& 0.279 & 8.64E1 & 7.35E1 & 0.220 \\
$^{ 75}$As & 100.00 & 198.6 & 8.8E-10 & M1+E2 & 0.389 & 9.81E-1 & 3.06 
& 3 & 0.310 & 7.64E1 & 7.5E1 & 0.242 \\
$^{ 193}$Ir & 62.70 & 73.0 & 6.1E-9 & M1+E2 & -0.558 & 1.43E-1 & 1.66E1 
& 2.37 & 0.670 & 4.35E2 & 6.2E1 & 0.564 \\
$^{ 153}$Eu & 52.20 & 151.6 & 3.6E-10 & (E1) &  & 3.35E-1 & 6.92 & 2.32 
& 0.572 & 1.6E1 & 5.37 & 0.462 \\
$^{ 141}$Pr & 100.00 & 145.4 & 1.9E-9 & M1+E2 & 0.069 & 6.84E-1 & 3.36 
& 2.3 & 0.533 & 1.98E1 & 1.36E1 & 0.433 \\
$^{ 189}$Os & 16.10 & 36.2 & 5.3E-10 & M1+E2 & 0.045 & 4.51E-2 & 4.82E1 
& 2.18 & 0.651 & 5.13E2 & 2.32E1 & 0.557 \\
$^{ 133}$Cs & 100.00 & 81.0 & 6.3E-9 & M1+E2 & -0.151 & 3.65E-1 & 5.34 
& 1.95 & 0.480 & 9.5E1 & 3.46E1 & 0.403 \\
$^{ 195}$Pt & 33.80 & 199.5 & 6.6E-10 & M1+E2 & 1.2 & 1.82E-1 & 9.78 & 1.78 
& 0.733 & 3.22E2 & 5.87E1 & 0.572 \\
$^{ 121}$Sb & 57.36 & 37.1 & 3.5E-9 & M1 &  & 8.26E-2 & 2.09E1 & 1.72 
& 0.437 & 4.05E1 & 3.34 & 0.374 \\
$^{ 149}$Sm & 13.80 & 22.5 & 7.1E-9 & M1+E2 & 0.0715 & 3.29E-2 & 5.21E1 
& 1.71 & 0.529 & 1.03E3 & 3.37E1 & 0.454 \\
$^{ 83}$Kr & 11.50 & 9.4 & 1.5E-7 & M1+E2 & 0.013 & 6E-2 & 2.85E1 & 1.71 
& 0.305 & 2.8E2 & 1.68E1 & 0.264 \\
$^{ 161}$Dy & 18.90 & 103.1 & 6E-10 & E1 &  & 1.78E-1 & 8.93 & 1.59 & 0.583 
& 1.75E1 & 3.12 & 0.484 \\
$^{ 201}$Hg & 13.18 & 32.2 & 1E-10 & M1+E2 & 0.017 & 1.08E-2 & 1.25E2 & 1.35 
& 0.684 & 5.13E2 & 5.54 & 0.586 \\
$^{ 145}$Nd & 8.30 & 67.2 & 2.9E-8 & E2 &  & 9.41E-2 & 1.41E1 & 1.33 & 0.521 
& 3.79E2 & 3.57E1 & 0.440 \\
$^{ 235}$U & 0.72 & 103.0 & 3.3E-11 & E2 &  & 5.36E-3 & 2.37E2 & 1.27 & 0.812 
& 6.75E3 & 3.61E1 & 0.674 \\
$^{ 191}$Ir & 37.30 & 82.4 & 4.1E-9 & M1+E2 & -0.88 & 8.34E-2 & 1.41E1 & 1.17 
& 0.673 & 3.8E2 & 3.17E1 & 0.564 \\
$^{ 139}$La & 99.91 & 165.9 & 1.5E-9 & M1 &  & 7.89E-1 & 1.48 & 1.17 & 0.522 
& 4.47 & 3.53 & 0.418 \\
$^{ 133}$Cs & 100.00 & 160.6 & 1.9E-10 & M1+E2 & 0.96 & 7.54E-2 & 1.39E1 
& 1.05 & 0.502 & 4.23E2 & 3.19E1 & 0.403 \\
$^{ 19}$F & 100.00 & 197.1 & 8.9E-8 & E2(calc) &  & 9.99E-1 & 1.03 & 1.03 
& 0.084 & 3.51E1 & 3.51E1 & 0.066 \\
$^{ 151}$Eu & 47.80 & 21.5 & 9.6E-9 & M1+E2 & 0.029 & 3.37E-2 & 2.95E1 
& 9.94E-1 & 0.537 & 3.13E2 & 1.06E1 & 0.462 \\
$^{ 195}$Pt & 33.80 & 129.7 & 6.7E-10 & E2 &  & 2.77E-2 & 3.59E1 & 9.94E-1 
& 0.699 & 1.07E3 & 2.97E1 & 0.572 \\
$^{ 201}$Hg & 13.18 & 26.3 & 6.3E-10 & M1+E2 & 0.02 & 1.27E-2 & 7.38E1 
& 9.39E-1 & 0.683 & 4.28E2 & 5.44 & 0.586 \\
$^{ 57}$Fe & 2.20 & 136.5 & 8.7E-9 & E2 &  & 1.04E-1 & 7.92 & 8.22E-1 
& 0.234 & 2.4E2 & 2.49E1 & 0.191 \\
$^{ 161}$Dy & 18.90 & 100.4 & 8.3E-10 & (E2) &  & 1.88E-2 & 4.19E1 & 7.89E-1 
& 0.582 & 1.19E3 & 2.24E1 & 0.484 \\
$^{ 187}$Os & 1.60 & 9.7 & 2.4E-9 & M1(+E2) & 0 & 4.31E-3 & 1.79E2 & 7.73E-1 
& 0.645 & 2.99E2 & 1.29 & 0.557 \\
$^{ 191}$Ir & 37.30 & 179.0 & 3.9E-11 & M1+E2 & -0.75 & 3.5E-2 & 1.84E1 
& 6.43E-1 & 0.712 & 5.56E2 & 1.94E1 & 0.564 \\
$^{ 189}$Os & 16.10 & 95.3 & 2.3E-10 & M1+E2 & 0.3 & 3.17E-2 & 2.01E1 
& 6.38E-1 & 0.668 & 4.76E2 & 1.51E1 & 0.557 \\
$^{ 45}$Sc & 100.00 & 12.4 & 3.2E-7 & (M2) &  & 2.34E-3 & 2.26E2 & 5.28E-1 
& 0.178 & 9.18E3 & 2.15E1 & 0.154 \\
$^{ 193}$Ir & 62.70 & 180.1 & 5.9E-11 & M1+E2 & -0.48 & 4.17E-2 & 8.24 
& 3.44E-1 & 0.713 & 2.24E2 & 9.35 & 0.564 \\
$^{ 161}$Dy & 18.90 & 74.6 & 3.1E-9 & E1 &  & 1.17E-1 & 1.74 & 2.03E-1 
& 0.575 & 3.08 & 3.59E-1 & 0.484 \\
$^{ 181}$Ta & 99.99 & 6.2 & 6E-6 & E1 &  & 3.23E-2 & 1.62 & 5.21E-2 & 0.619 
& 1.99 & 6.43E-2 & 0.535 \\
$^{ 157}$Gd & 15.65 & 63.9 & 4.6E-7 & E1 &  & 2.48E-1 & 9.63E-2 & 2.39E-2 
& 0.555 & 1.64E-1 & 4.07E-2 & 0.469 \\
$^{ 67}$Zn & 4.10 & 93.3 & 9.2E-6 & E2 &  & 5.38E-1 & 3.18E-2 & 1.71E-2 
& 0.264 & 8.91E-1 & 4.79E-1 & 0.220 \\
$^{ 73}$Ge & 7.73 & 13.3 & 3E-6 & E2 &  & 1.01E-3 & 6.75 & 6.81E-3 & 0.272 
& 1.65E2 & 1.66E-1 & 0.235 \\
$^{ 153}$Eu & 52.20 & 172.9 & 1.4E-10 & M1+E2 & 0.77 & 2.43E-3 & 4.39E-1 
& 1.07E-3 & 0.580 & 1.32E1 & 3.21E-2 & 0.462 \\
$^{ 153}$Eu & 52.20 & 103.2 & 3.9E-9 & M1+E2 & 0.127 & 1.18E-2 & 7.21E-2 
& 8.49E-4 & 0.556 & 9.82E-1 & 1.16E-2 & 0.462 \\
$^{ 161}$Dy & 18.90 & 131.8 & 1.4E-10 & E1(calc) &  & 1.37E-3 & 8.82E-2 
& 1.21E-4 & 0.592 & 1.91E-1 & 2.62E-4 & 0.484 \\
$^{ 171}$Yb & 14.30 & 122.4 & 2.7E-7 & E2(calc) &  & 1.29E-3 & 5.73E-3 
& 7.38E-6 & 0.624 & 1.69E-1 & 2.18E-4 & 0.513 \\
$^{ 77}$Se & 7.63 & 161.9 & 1.7E1 & E3 &  & 5.29E-1 & 4.63E-7 & 2.45E-7 
& 0.311 & 5.8E-4 & 3.07E-4 & 0.249 \\
$^{ 151}$Eu & 47.80 & 196.0 & 5.9E-5 & E3(calc) &  & 1.29E-3 & 3.88E-5 
& 5E-8 & 0.590 & 5.38E-2 & 6.94E-5 & 0.462 \\
$^{ 109}$Ag & 48.16 & 88.0 & 4E1 & E3 &  & 3.58E-2 & 1.26E-6 & 4.5E-8 
& 0.412 & 1.27E-3 & 4.56E-5 & 0.345 \\
$^{ 107}$Ag & 51.84 & 93.1 & 4.4E1 & E3 &  & 4.64E-2 & 9.67E-7 & 4.48E-8 
& 0.413 & 9.93E-4 & 4.61E-5 & 0.345 \\
$^{ 103}$Rh & 100.00 & 39.8 & 3.4E3 & E3 &  & 6.87E-4 & 8.72E-8 & 5.99E-11 
& 0.386 & 7.71E-5 & 5.29E-8 & 0.330 \\
$^{ 83}$Kr & 11.50 & 32.1 & 6.6E3 & E3 &  & 4.97E-4 & 7.31E-9 & 3.63E-12 
& 0.308 & 6.33E-6 & 3.15E-9 & 0.264 \\
\hline
\end{tabular}

\end{document}